\documentclass[twocolumn,showpacs,amsmath,amssymb,prl]{revtex4} 
 
\usepackage{graphicx}   
\usepackage{dcolumn}    
\usepackage{bm}         
\newcommand{\rd}{{\rm d}} 
\newcommand{\re}{{\rm e}} 
\newcommand{\ri}{{\rm i}} 
 
\newcommand{\Jz}{{\rm J}_0} 
 
\begin{document}

\title[Superfluid-insulator transition]{Superfluid-insulator transition  
       in a periodically driven optical lattice}  
 
\author{Andr\'e Eckardt} 
\author{Christoph Weiss} 
\author{Martin Holthaus} 
   
\affiliation{Institut f\"ur Physik, Carl von Ossietzky Universit\"at, 
        D-26111 Oldenburg, Germany} 
                  
\date{October 23, 2005} 
 
\begin{abstract} 
We demonstrate that the transition from a superfluid to a Mott insulator 
in the Bose--Hubbard model can be induced by an oscillating force through 
an effective renormalization of the tunneling matrix element. The  
mechanism involves adiabatic following of Floquet states, and can be 
tested experimentally with Bose--Einstein condensates in periodically  
driven optical lattices. Its extension from small to very large systems  
yields nontrivial information on the condensate dynamics.  
\end{abstract} 
 
 
\maketitle 
 
                 
The Bose--Hubbard model plays an important role in condensed matter physics, 
since it embodies essential features of strongly interacting Bose systems 
in a minimal manner, namely the competition between kinetic and 
potential energy effects, and the resulting quantum phase transition from  
a superfluid to a Mott insulator~\cite{FisherEtAl89,Sachdev99}. 
It describes Bose particles on a lattice with on-site interaction, so that  
particles occupying the same lattice site repel each other, while tunneling  
is allowed between adjacent sites. This is expressed by the Hamiltonian  
\begin{equation} 
        \hat{H}_0 = -J\sum_{\langle i,j \rangle} 
        \left( \hat{c}^{\dag}_i \hat{c}_j + \hat{c}^{\dag}_j \hat{c}_i \right) 
        + \frac{U}{2} \sum_j \hat{n}_j ( \hat{n}_j - 1 ) \; ,  
\label{eq:HBH} 
\end{equation}    
where $\hat{c}^{(\dag)}_j$ is an annihilation (creation) operator for a  
boson on the site labeled~$j$, and $\hat{n}_j = \hat{c}^{\dag}_j \hat{c}_j$ 
denotes the corresponding number operator. The first sum runs over all 
pairs of neighboring sites $i$ and $j$, with the matrix element~$J$  
quantifying the strength of the tunneling contact. Moreover, $U$  
is the repulsion energy contributed by one pair of bosons located on the  
same site. Therefore, the characteristic  dimensionless parameter is the  
ratio $U/J$: When $U \ll J$, so that tunneling dominates, the ground state  
of the system describes a superfluid, whereas it has the properties of a Mott  
insulator when the interaction dominates, $U \gg J$. Mean-field  
theory~\cite{FisherEtAl89,Sachdev99} gives the critical value  
$(U/J)_c \approx z \times 5.83$ for the transition in a lattice filled  
with one particle per site, which captures the case of a three-dimensional  
($3d$) cubic lattice with coordination number $z = 6$ reasonably well, whereas  
more refined methods~\cite{ElstnerMonien99} yield a Kosterlitz-Thouless  
transition with $(U/J)_c \approx 3.8$ for $d = 1$. 
 
After the model~(\ref{eq:HBH}) had long been of primarily theoretical  
interest, it has found its laboratory realization with Bose--Einstein  
condensates in optical lattices~\cite{JakschEtAl98,Zwerger03,JakschZoller05}.  
In such systems, the expected transition has been observed upon varying the 
lattice depth, both for $d = 3$~\cite{GreinerEtAl02} and  
$d = 1$~\cite{StoeferleEtAl04}.  In this Letter, we demonstrate that the  
transition from a superfluid to a Mott insulator can be induced in an  
altogether different manner which, in contrast to all scenarios studied before, 
hinges on the effect of a time-dependent force, and which can be assessed  
experimentally with condensates in periodically modulated optical lattices.  
We will investigate a {\em periodically forced\/} Bose--Hubbard model for  
$d = 1$, as described by the explicitly time-dependent Hamiltonian 
\begin{equation} 
        \hat{H}(t) = \hat{H}_0 + K \cos(\omega t) \sum_j j \, \hat{n}_j \; , 
\label{eq:DBH} 
\end{equation}   
where the equidistant sites are labeled according to their position in  
ascending order. The oscillating term, which mimics a monochromatic electric  
dipole potential with frequency~$\omega$ and amplitude~$K$, can be realized  
experimentally by periodically shifting the position of a mirror employed to  
generate the standing laser wave, and transforming to the co-moving frame of  
reference~\cite{GrahamEtAl92,MadisonEtAl98}. We will argue that the driven  
system~(\ref{eq:DBH}) behaves, for sufficiently high frequencies, similar  
as the undriven system~(\ref{eq:HBH}), but with the tunneling matrix  
element~$J$ of the latter being replaced by the effective matrix element   
\begin{equation} 
        J_{\rm eff} = J \, \Jz\big(K/(\hbar\omega)\big) \; , 
\label{eq:REN} 
\end{equation}      
where $\Jz(x)$ denotes the ordinary Bessel function of order zero. Hence, 
the actual control parameter becomes $U/J_{\rm eff}$, which can be varied  
by adjusting the parameters of the periodic modulation. This implies the 
possibility to switch between the superfluid and the insulator state by  
changing, {\em e.g.\/}, the modulation strength~$K$.   
 
The rescaling~(\ref{eq:REN}) is not unfamiliar with periodically driven 
single-particle quantum systems. It occurs, among others, when a particle    
moves on a periodically forced $1d$~lattice with nearest neighbor  
coupling~\cite{DunlapKenkre86,Holthaus92}, such as an electron in a 
semiconductor superlattice~\cite{MeierEtAl95}. It also underlies the 
$\Jz$-type renormalization of atomic $g$-factors in oscillating magnetic  
fields~\cite{HarocheEtAl70,CohenTannoudji94}, and the coherent destruction  
of tunneling of a particle in a periodically forced  
double-well~\cite{GrossmannEtAl91,GrifoniHanggi98}. However, as will be 
discussed below, the many-body system~(\ref{eq:DBH}) is significantly more  
involved when the thermodynamic limit is taken; the rescaling~(\ref{eq:REN})  
then describes only part of the relevant physics.        
 
Our analysis is based on quantum Floquet theory~\cite{Shirley65}: Since  
the Hamiltonian~(\ref{eq:DBH}) depends periodically on time,  
$\hat{H}(t) = \hat{H}(t+ T)$ with period $T = 2\pi/\omega$, there exists  
a complete set of solutions to the time-dependent many-body Schr\"odinger  
equation of the form  
$|\psi_n(t)\rangle = |u_n(t)\rangle \exp(-\ri\varepsilon_n t/\hbar)$, 
where the Floquet functions $|u_n(t)\rangle$ inherit the period of the 
driving force, satisfying $|u_n(t)\rangle = |u_n(t+T)\rangle$. Thus, 
Floquet states for periodically time-dependent quantum systems, obtained 
by solving the eigenvalue equation 
\begin{equation} 
        \left( \hat{H}(t) - \ri\hbar\partial_t \right) |u_n(t)\rangle 
        = \varepsilon_n |u_n(t)\rangle \; , 
\label{eq:EIG} 
\end{equation} 
constitute an analog of Bloch states known from spatially periodic crystals;  
the eigenvalues $\varepsilon_n$, which describe the time evolution of these  
states in close analogy to the evolution of energy eigenstates, are called  
quasienergies. While in solid-state physics quasimomenta are defined up to  
an integer multiple of a reciprocal lattice vector, quasienergies are  
defined up to an integer multiple of $\hbar\omega$: If $|u_n(t)\rangle$  
solves Eq.~(\ref{eq:EIG}) with eigenvalue $\varepsilon_n$, 
and $ m = 0, \pm 1, \pm 2, \ldots\,$, then $|u_n(t)\rangle\exp(\ri m \omega t)$  
is a $T$-periodic eigensolution with quasienergy $\varepsilon_n + m\hbar\omega$. 
The quasienergy spectrum of a periodically time-dependent quantum system thus  
possesses a Brillouin zone-like structure, the width of one zone being  
$\hbar\omega$. 
 
We then employ the Floquet basis 
\begin{equation} 
        |\{n_j\},m \rangle = |\{n_j\}\rangle  
        \exp\Big[\! -\ri\frac{K}{\hbar\omega}\sin(\omega t) \sum_j j n_j 
                    +\ri m \omega t \Big] \; , 
\label{eq:BAS} 
\end{equation} 
where $|\{n_j\}\rangle$ indicates a Fock state with $n_j$ particles on 
the $j$th site, and $m$ again accounts for the zone structure. The  
eigenvalue problem~(\ref{eq:EIG}) refers to an extended Hilbert space of  
$T$-periodic functions, in which the time variable is regarded as a  
coordinate~\cite{Sambe73}, so that the scalar product in that space is 
given by 
\begin{equation} 
        \langle \langle \cdot | \cdot \rangle\rangle 
        = \frac{1}{T} \int_0^T \! \rd t \,  
        \langle \cdot | \cdot \rangle \; ,  
\label{eq:SCP} 
\end{equation}    
{\em i.e.\/}, by the usual scalar product $\langle \cdot | \cdot \rangle$ 
combined with time-averaging. Hence, the quasienergies are obtained by 
computing the matrix elements of the operator  
$\hat{H}(t) - \ri\hbar\partial_t$ in the basis~(\ref{eq:BAS}) with respect 
to the scalar product~(\ref{eq:SCP}), and diagonalizing. Denoting the 
Hamiltonian~$\hat{H}_0$ with $J = 0$, which is diagonal in the 
basis~(\ref{eq:BAS}), by $\hat{H}_{\rm int}$, and its $J$-proportional 
tunneling term by $\hat{H}_{\rm tun}$, we find 
\begin{eqnarray} 
        & &  
        \langle\langle \{n_j'\},m' | \hat{H}(t) - \ri\hbar\partial_t | 
        \{n_j\},m \rangle\rangle  
\nonumber \\  
        & = & 
        \delta_{m',m}\left[\langle \{n_j'\} | \hat{H}_{\rm int} | 
        \{n_j\} \rangle + m\hbar\omega\right] 
\nonumber \\ 
        & + & 
        s^{m'-m} {\rm J}_{m'-m}\big(K/(\hbar\omega)\big) \, 
        \langle \{n_j'\} | \hat{H}_{\rm tun} | \{n_j\} \rangle \; , 
\end{eqnarray} 
where $s = \sum_j (n_j' - n_j)j = \pm 1$, since $\hat{H}_{\rm tun}$ only  
transfers one particle by one site. We observe that with respect to the 
``photon'' index $m$ this matrix has a transparent block structure: 
The diagonal blocks with $m = m'$ reproduce the matrix which yields the 
eigenvalues of the undriven system, but with $J$ replaced by $J_{\rm eff}$ 
according to Eq.~(\ref{eq:REN}), and replicas shifted by integer multiples 
of $\hbar\omega$. These blocks are coupled by nondiagonal ones proportional 
to Bessel functions ${\rm J}_{m' - m}\big(K/(\hbar\omega)\big)$. Obviously,  
the anticipated rescaling~(\ref{eq:REN}) holds only to the extent that these  
couplings can be neglected. This will be the case, at least in a perturbative  
sense, if the block separation $\hbar\omega$ is much larger than both the  
energy scale~$J$ of the coupling and the energy scale~$U$ associated with  
the diagonal blocks, {\em i.e.\/}, for high frequencies  
$\hbar\omega \gg \max\{J,U\}$.

\begin{figure}[t] 
\includegraphics[width = 0.8\linewidth]{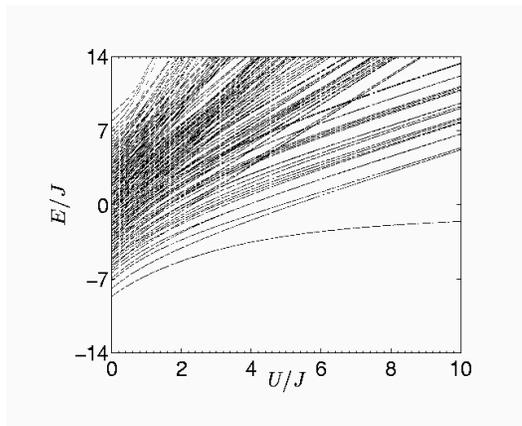}  
\caption{Exact energy spectrum of a small one-dimensional undriven  
        lattice~(\ref{eq:HBH}) with $M = 5$ sites and $N = 5$ particles,  
        versus Mott-Hubbard parameter $U/J$. The splitting-off of the ground  
        state with increasing interaction strength can be regarded as a  
        precursor of the quantum phase transition.} 
\label{F_1} 
\end{figure}

To demonstrate that this reasoning is justified, we present numerical results  
for small systems. Figure~\ref{F_1} depicts the exact energy spectrum for a  
one-dimensional undriven model~(\ref{eq:HBH}) with $N = 5$ particles on  
$M = 5$ sites. Even here, the precursor of the superfluid-insulator transition  
already is apparent: With increasing~$U/J$ the system's ground state,  
associated with a uniform distribution of the particles over the sites, splits  
off from the group of excited states, which describe various patterns of  
particle-hole excitations. In the limit of an infinitely large system,  
$N \to \infty$ and $M \to \infty$ with $N/M = 1$ held constant, the excited  
states form continuous energy bands; the ground state then splits off from  
the lowest band at a finite $(U/J)_c$~\cite{ElstnerMonien99}. This separation  
of an individual state from the continuum indicates the transition to the  
Mott insulator state.

\begin{figure}[t] 
\includegraphics[width=0.8\linewidth]{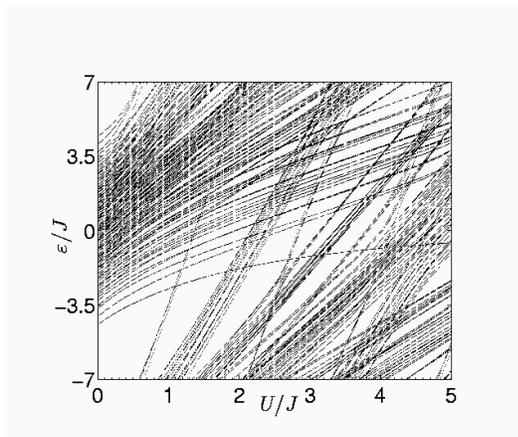} 
\caption{First Brillouin zone of the exact quasienergy spectrum of the driven  
        $1d$ system~(\ref{eq:DBH}) with $N = M = 5$, $\hbar\omega/J = 14$, and 
        $K/\hbar\omega = 1.5$. This spectrum is recovered approximately from  
        the energy eigenvalues shown in Fig.~\ref{F_1}, if $J$ is replaced  
        by $J_{\rm eff} \approx J/2$ here, and the energies are taken modulo  
        $\hbar\omega$.} 
\label{F_2} 
\end{figure}

For comparison, Fig.~\ref{F_2} shows the first Brillouin zone of numerically  
computed quasienergies for the driven system~(\ref{eq:DBH}), again with  
$N = M = 5$, scaled frequency $\hbar\omega/J = 14$, and scaled driving  
amplitude $K/\hbar\omega = 1.5$. Since $\Jz(1.5) \approx 0.5$, this set of  
parameters allows for a convenient test of the hypothesis~(\ref{eq:REN}):  
With $J_{\rm eff} \approx J/2$, the {\em quasi\/}energy spectrum of the  
driven system~(\ref{eq:DBH}) for a given parameter $U/J$ should correspond  
(apart from its zone structure) to the {\em energy\/} spectrum of the  
undriven system~(\ref{eq:HBH}) with the same $U/J_{\rm eff}$, which is  
about $2U/J$. This is borne out, to remarkable accuracy, by a comparison  
of Figs.~\ref{F_1} and~\ref{F_2}: With the scales of the respective axes  
differing by a factor of two, the eigenvalues plotted in Fig.~2 almost  
equal those in Fig.~1. As a consequence of the Brillouin zone structure,  
quasienergy eigenvalues which disappear at the upper zone boundary reappear  
again at the lower one. This reappearance is a source of substantial  
complications: States originating from different Brillouin zones are  
coupled through the matrix elements neglected in the explanation of the  
renormalization~(\ref{eq:REN}), so that many apparent level crossings in  
Fig.~\ref{F_2} actually are tiny avoided crossings. It is the high-frequency  
condition $\hbar\omega \gg \max\{J,U\}$ which guarantees that these avoided  
crossings remain too narrow to be resolved. If this condition is not met,  
a multitude of large avoided crossings appears in the spectrum, thus  
revealing typical signatures of quantum chaos~\cite{Haake04}.

\begin{figure}[t] 
\includegraphics[width=0.8\linewidth]{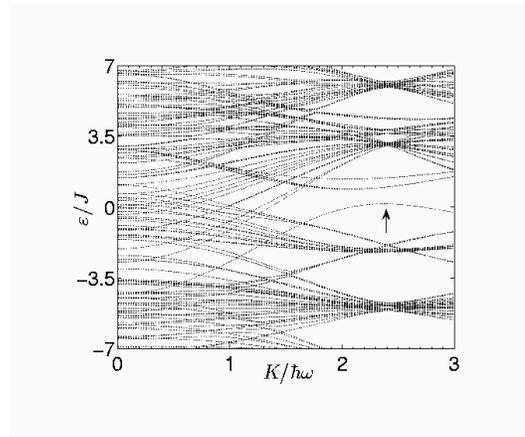} 
\caption{First Brillouin zone of the exact quasienergy spectrum of the  
        driven $1d$ system with $N = M = 5$, $\hbar\omega/J = 14$, and  
        $U/J = 3$, versus scaled driving amplitude $K/\hbar\omega$. For  
        $K/\hbar\omega \approx 2.4$, close to the first zero of $\Jz$, the  
        tunneling contact is (almost) switched off, resulting in a collapse  
        of the different bands. In the vicinity of this value, the Floquet  
        state evolving from the system's ground state (marked by the arrow) has  
        the properties of a Mott insulator, even though the undriven system's  
        ground state is superfluid for $U/J = 3$.} 
\label{F_3} 
\end{figure}

Figure~\ref{F_3} shows quasienergies for $U/J = 3$ kept fixed, again for 
$\hbar\omega/J = 14$, as functions of the scaled amplitude $K/(\hbar\omega)$.  
For $K/(\hbar\omega) \approx 2.4$, close to the first zero of $\Jz$, the 
tunneling contact is quenched almost entirely, so that the various bands of 
particle-hole excitations collapse. In the vicinity of this point, the Floquet  
state evolving from the unperturbed ground state must have the properties of 
a Mott insulator, although the ground state of the undriven system describes  
a superfluid for $U/J = 3$.

\begin{figure}[t] 
\includegraphics[width=0.8\linewidth]{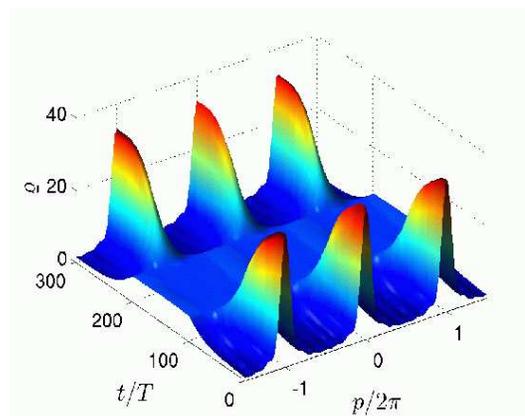} 
\caption{Time evolution of the momentum distribution $\varrho(p)$ obtained  
        by solving the time-dependent Schr\"odinger equation for $N = M = 7$,  
        $\hbar\omega/J = 14$, and a slowly varying amplitude $K/(\hbar\omega)$  
        which is linearly ramped up to $2.4$ during the first $100$ cycles~$T$, 
        then stays constant for another $100 \, T$, and is ramped down to zero  
        during the final $100$ cycles. The system was initially in its ground  
        state; the distribution was recorded at integer multiples of~$T$.  
        The disappearance and  reappearance of the peak pattern signals the  
        transition from the superfluid to the Mott state and back.}    
\label{F_4} 
\end{figure}

To verify this conclusion, we have solved the time-dependent Schr\"odinger 
equation for a system with $N = M =7$ and $U/J = 3$, initially prepared 
in its superfluid ground state and then subjected to periodic forcing with  
frequency $\hbar\omega/J = 14$, and an amplitude $K/(\hbar\omega)$ which  
increases linearly from $0$ to $2.4$ during the first $100$ cycles~$T$, 
then stays constant for another $100$~cycles, and finally is linearly ramped  
down to zero between $t = 200 \, T$ and $t = 300 \, T$. Experimentally,  
the superfluid phase is detected by a sharply peaked reciprocal lattice  
pattern in the momentum distribution,  
$\varrho(p) = \sum_{j,k}\langle{c_j^\dagger c_k\rangle\re^{-\ri p (j - k)}}$,  
which can be measured by time-of-flight absorption imaging. (Here, $p$ is 
given in multiples of $\hbar$/lattice constant.) Figure~\ref{F_4} depicts 
our result: Since Floquet states respect an approximate adiabatic  
principle~\cite{BreuerHolthaus89}, the initial superfluid ground state is  
first adiabatically transformed into a Mott insulator state, as witnessed by  
the disappearance of the peaked momentum distribution, and then transformed  
back to the inital state, apart from a remaining excitation of other 
states totalling to a few percent. This is a major result: The amplitude of  
the periodic force decides whether the system is superfluid, or in a Mott  
insulator state. 
    
When extrapolating from these model calculations to large systems, two issues  
have to be considered. Firstly, when increasing the number of lattices sites  
while maintaining an occupancy of one particle per site, say, the quasienergy  
levels fill the Brillouin zone densely; in the thermodynamic limit, the  
spectrum probably is a continuum. Then there will be no ``sharp'' Floquet  
state evolving from the ground state, but rather a resonance with a finite  
lifetime, due to the residual couplings to other states. Starting from an  
undriven, infinite system with a superfluid ground state, and switching on  
the periodic force, we conjecture that a Mott-insulator-like resonance appears  
at that amplitude~$K$ which, after rescaling according to Eq.~(\ref{eq:REN}),  
corresponds to that tunneling matrix element~$J$ which marks the  
quantum phase transition in the undriven system. Secondly, for an infinitely  
large lattice there is no adiabatic limit when switching on the driving  
force~\cite{HoneEtAl97}; turn-on and turn-off necessarily have to take place  
within a short interval. Our calculations indicate that adiabatic following  
can even be improved by {\em shortening\/} the turn-on time, since  
Landau-Zener transitions at narrow avoided crossings, possibly corresponding  
to condensate heating, then are suppressed.      
  
While these issues are still not covered by rigorous mathematical theorems  
on periodically driven quantum systems~\cite{Howland92}, and remain out of  
reach of even most powerful supercomputers, they can be addressed in the  
laboratory. In experiments with cold atoms in driven optical  
lattices~\cite{MadisonEtAl98}, narrowing of Bloch bands compatible with  
the rescaling~(\ref{eq:REN}) has already been observed, even though the  
single-band regime has not been reached. Employing Bose--Einstein  
condensates in optical lattices in order to realize the driven Bose--Hubbard  
model~(\ref{eq:DBH}), one has to respect not only the high-frequency condition 
$\hbar\omega > \max\{J,U\}$ required for the approximate Bessel-function 
rescaling~(\ref{eq:REN}), but there is the obvious additional condition 
$\hbar\omega < \Delta$, where $\Delta$ denotes the gap between the lowest  
two Bloch bands of the undriven lattice, in order to exclude transitions  
to higher band states. Elementary estimates in the spirit of  
Ref.~\cite{Zwerger03} suggest that under typical conditions (as provided by  
$^{87}$Rb atoms in a lattice created by laser radiation of $\lambda = 852$~nm  
wavelength~\cite{GreinerEtAl02}) this leaves a viable window of frequencies  
in the low kHz regime. For higher filling factors, or in $3d$~lattices with  
forcing in all three directions, the critical parameter $(U/J)_c$ becomes much  
larger, allowing one to employ deeper lattices with larger band gap~$\Delta$,  
and hence to work with still higher frequencies without violating the  
single-band approximation. A quantity of key interest in such experiments  
will be the extent to which, after starting from a superfluid ground state,  
then ramping up the force into the insulator regime and ramping it down again  
as in Fig.~\ref{F_4}, the superfluid peak pattern reappears, providing  
information on both the lifetime of the conjectured Mott-like resonance state  
and the degree of adiabatic following, or, more generally, on the extent to  
which the quantum evolution of a mesoscopic matter wave can be guided even  
under critical conditions. Thus, the scenario envisioned here is not intended  
as a look at the common superfluid-insulator transition from a different  
angle, but aims at obtaining genuinely new, nontrivial information on  
condensate dynamics.  
     
This work was supported by the DFG through the Priority Programme SPP~1116.  
A.E.\ acknowledges a fellowship from the Studienstiftung des deutschen Volkes.

\end{document}